\documentclass[preprint,superscriptaddress,amsmath]{revtex4-1}
\pdfoutput=1
\usepackage{graphicx}
\usepackage{slashed}
\usepackage{bm}
\usepackage[T1]{fontenc} 
\usepackage{setspace}
\usepackage{float}
\usepackage{verbatim}
\def\be{\begin{equation}}
\def\ee{\end{equation}}
\def\bea{\begin{array}}
	\def\eea{\end{array}}
\def\beqa{\begin{eqnarray}}
\def\eeqa{\end{eqnarray}}
\def\beqas{\begin{eqnarray*}}
	
	\def\eeqas{\end{eqnarray*}}

\def\bp{\begin{picture}}
\def\ep{\end{picture}}
\def\bc{\begin{center}}
	\def\ec{\end{center}}
\def\bfig{\begin{figure}}
	\def\efig{\end{figure}}

\def\bit{\begin{itemize}}
	\def\eit{\end{itemize}}

\def\[{\left[}
\def\]{\right]}
\def\({\left(}
\def\){\right)}

\def\..{\left.}
\def\.{\right.}

\def\ep{\epsilon}

\def\max{\mathop{\rm max}}

\begin{document}


\title{Testing electroweak SUSY for muon $g-2$ and dark matter at the LHC and beyond}
\author{Murat Abdughani}
\affiliation{CAS Key Laboratory of Theoretical Physics, Institute of Theoretical
                Physics, Chinese Academy of Sciences, Beijing 100190, P. R. China}
\affiliation{School of Physical Sciences, University of Chinese Academy of Sciences,\\ Beijing 100049, P. R. China}
\affiliation{Department of Physics and Institute of Theoretical Physics, Nanjing Normal University, Nanjing 210023, P. R. China}

\author{Ken-ichi Hikasa}
\affiliation{Department of Physics, Tohoku University, Sendai 980-8578, Japan}

\author{Lei Wu}
\affiliation{Department of Physics and Institute of Theoretical Physics, Nanjing Normal University, Nanjing 210023, P. R. China}

\author{Jin Min Yang}
\affiliation{CAS Key Laboratory of Theoretical Physics, Institute of Theoretical
                Physics, Chinese Academy of Sciences, Beijing 100190, P. R. China}
\affiliation{School of Physical Sciences, University of Chinese Academy of Sciences,\\ Beijing 100049, P. R. China}

\author{Jun Zhao}
\affiliation{CAS Key Laboratory of Theoretical Physics, Institute of Theoretical
                Physics, Chinese Academy of Sciences, Beijing 100190, P. R. China}
\affiliation{School of Physical Sciences, University of Chinese Academy of Sciences,\\ Beijing 100049, P. R. China}

\begin{abstract}	
Given that the LHC experiment has produced strong constraints on the colored supersymmetric particles (sparticles),
testing the electroweak supersymmetry (EWSUSY) will be the next crucial task at the LHC. On the other hand,
the light electroweakinos and sleptons in the EWSUSY can also contribute to the dark matter (DM) and low energy
lepton observables. The precision measurements of them will provide the indirect evidence of SUSY. In this work,
we confront the EWSUSY with the muon $g-2$ anomaly, the DM relic density, the direct detection limits and the
latest LHC Run-2 data. We find that the sneutrino DM or the neutralino DM with sizable higgsino component
has been excluded by the direct detections. Then two viable scenarios are pinned down: one has the
light compressed bino and sleptons but heavy higgsinos, and the other has the light compressed bino, winos
and sleptons. In the former case, the LSP and slepton masses have to be smaller than about 350 GeV.
While in the latter case, the LSP and slepton masses have to be smaller than about 700 GeV and 800 GeV,
respectively. From investigating the observability of these sparticles in both scenarios at future colliders,
it turns out that the HE-LHC with a luminosity of 15 ab$^{-1}$ can exclude the whole BHL and most part of BWL scenarios at $2\sigma$ level. The precision measurement of the Higgs couplings at the lepton colliders
could play a complementary role of probing the BWL scenario.
\end{abstract}

\maketitle
	
\section{Introduction}\label{sec:intro}
In particle physics, the muon anomalous magnetic moment $a_\mu$ is one of the most precisely measured quantities.
Since the first results were reported, there has been a longstanding $\sim 3\sigma$ discrepancy between theory
and experiment, which triggered numerous studies of new physics explanations. As a successor of the previous
E821 experiment performed at BNL, the on-going muon $g-2$ experiment E989 at Fermilab is to measure $a_\mu$ with
a relative precision of 140 parts-per-billion (ppb)~\cite{Driutti:2018ynq}. This precision is a factor of four
improvement from the current experiment~\cite{gm2-value}. If this anomaly still persists, it would be a
clear evidence for new physics beyond the Standard Model (BSM).

Meanwhile, dark matter (DM) that constitutes the majority of matter in the universe has been established by
astrophysical and cosmological observations. Understanding its nature and interactions is one of the most
important quests of contemporary physics. The Weakly Interacting Massive Particle (WIMP) paradigm provides
an attractive solution to the DM issue as it can naturally produce the measured relic density
through the robust mechanism of thermal freeze-out. Therefore, various concrete realizations of WIMP models
have been proposed, which has been being tested in DM (in)direct detections and collider experiments~\cite{Arcadi:2017kky}.

Among new physics models for solving these two problems, supersymmetry (SUSY) is one of the most popular
candidates, which has a beautiful mathematical structure and is considered as a part of a larger vision
of physics. In supersymmetric models, the lightest neutralino $\tilde{\chi}^0_1$ can serve as a natural
WIMP DM candidate if the $R$-parity is conserved. Meanwhile, the muon $g-2$ anomaly can be explained
by the contributions of light sleptons and electroweakinos running in the loops~\cite{g2-mssm-0,g2-mssm-2,g2-mssm-3,g2-mssm-31}. In addition, SUSY
can also solve the hierarchy problem and realize the unification of gauge couplings at the GUT scale.
Due to its overwhelming virtues and popularity, the low energy SUSY has long been pursued by both
theorists and experimentalists.

Up to now, the LHC null observation of colored sparticles has excluded the masses of squarks and gluinos
lighter than about 2 TeV in simplified models~\cite{ATLAS:2019vcq}. Fortunately, to account for the DM abundance and the muon
$g-2$ anomaly, only the uncolored sparticles (electroweakinos and sleptons) need to be
light, which are subject to relatively rather weak constraints from the LHC searches~\cite{Athron:2018vxy}. As a result,
the electroweak SUSY that only consists of light electroweakinos and sleptons is strongly favored
by current experimental data. As shown in Refs.~\cite{Yanagida:2017dao,Wang:2017vxj,Wang:2015nra,Wang:2015rli,g2-mssm-15,g2-mssm-16,g2-mssm-17,g2-mssm-19,g2-mssm-22,g2-mssm-23,g2-mssm-24,g2-mssm-25,g2-mssm-26,g2-mssm-28,g2-mssm-29,Abe:2002eq,Endo:2011xq,Tran:2018kxv,Fukuyama:2016mqb}, such a scenario can be realized in well-motivated high scale supersymmetric models.

In this work we perform a comprehensive study of the phenomenology of the EWSUSY scenario for the muon $g-2$
and dark matter. Note that in the literature~\cite{Cox:2018qyi,Kobakhidze:2016mdx,g2-mssm-6,g2-mssm-10,g2-mssm-27,g2-mssm-30,Sabatta:2019nfg,Padley:2015uma}
such a scenario has been discussed to some extent. However, those studies either did not require the
SUSY dark matter to provide the correct abundance or focused on
the phenomenology at the LHC. Unlike them, we pin down the viable parameter space of EWSUSY
for explaining the dark matter abundance and the muon $g-2$ anomaly by a numerical scan. We find that
the masses of the electroweakinos and sleptons are bounded in certain ranges, which will guide the
search strategies at colliders. In addition to the LHC observability of such a scenario, we will also
explore its test at the LHC upgrades and the $e^+e^-$ Higgs factory. As a precison test machine,
the future $e^+e^-$ Higgs factories, such as CEPC, FCC-ee and ILC, have limited energy to directly
search for SUSY particles. However, they can test the low energy SUSY through the precision measurements
of Higgs couplings.

The structure of this work is organized as follows. In Sec.~II, we will recapitulate the studies of
muon $g-2$ and neutralino DM in the MSSM. In Sec.~III, we perform a numerical scan over the parameter space
of EWSUSY and discuss the implications for sparticles. In Sec.~IV, we investigate the prospects of hunting
for the electroweakinos and sleptons in those scenarios at the LHC and future colliders.
Finally, we draw our conclusions in Sec. V.

\section{neutralino dark matter and muon $g-2$ in the MSSM}
In the MSSM there are four neutralinos $\tilde{\chi}^0_{1,2,3,4}$ that are the mixtures
of bino ($\tilde{B}$), wino ($\tilde{W}^0$) and
neutral higgsinos ($\tilde{H}_{u,d}^0$).
The mass matrix is given by~\cite{Haber:1985rc}
\begin{eqnarray}
	M_{\tilde{\chi}^0} &=& \left(\begin{array}{cccc}
		M_1 & 0 & -c_{\beta} s_W m_Z & s_{\beta} s_W  m_Z \\
		0 & M_2 & c_{\beta} c_W m_Z & -s_{\beta} c_W m_Z  \\
		-c_{\beta} s_W m_Z & c_{\beta} c_W m_Z & 0 & -\mu  \\
		s_{\beta} s_W m_Z & -s_{\beta} c_W m_Z& -\mu & 0
	\end{array} \right)
	\label{neutralinomassmatrix}
\end{eqnarray}
where $s_{\beta}$, $c_{\beta}$, $s_W$ and  $c_W$ stands respectively
for $\sin\beta$, $\cos\beta$, $\sin\theta_W$ and $\cos\theta_W$. $M_1$ and $M_2$ are the soft-breaking mass parameters for bino and wino, respectively. $\mu$ is the higgsino mass parameter. We can diagonalize Eq.~(\ref{neutralinomassmatrix}) by a unitary $4\times 4$ matrix $N$. Besides, the mass matrix of charginos that are the mixtures of wino ($\tilde{W}^\pm$)
and charged higgsinos ($\tilde{H}_d^-$, $\tilde{H}_u^+$) can be written as
\begin{eqnarray}
	M_{\tilde{\chi}^\pm}&=&\left( \begin{array}{cc}
		M_2& \sqrt{2} s_{\beta} m_W \\
		\sqrt{2} c_{\beta} m_W& \mu
	\end{array} \right)
	\label{mass}
\end{eqnarray}
which can be diagonalized by two unitary $2\times 2$ matrices $U$ and $V$.

In the MSSM the lightest sparticle (LSP) can be the neutralino $\tilde{\chi}^0_1$, which can play the role of dark matter. It is a mixture of bino, wino and higgsinos.
Depending on its dominant component, the LSP $\tilde{\chi}^0_1$ can be bino-like, higgsino-like or wino-like. When $\tilde{\chi}^0_1$ is wino-like or higgsino-like, it usually has too large annihilation rates in the early universe to produce sufficient dark matter relic density. If we require them to
provide the correct dark matter abundance without other non-SUSY dark matter components like axions, the masses of higgsino-like and wino-like dark matter have to be at TeV scale~\cite{ArkaniHamed:2006mb,Han:2013usa}, which results in too heavy electroweakino spectrum to generate sizable contributions to muon $g-2$. On the other hand, the wino-like or higgsino-like dark matter scattering with nucleon has a sizable cross section and thus subject to stringent limits from dark matter direct detection experiments. Besides, it should be noted that the sneutrino in our study can be dark matter as well, which, however, was excluded by the direct detection. Therefore, in our study, we will focus on the bino-like dark matter, which can give the observed relic density by mixing with higgsino/wino, resonantly annihilating through $Z$/Higgs bosons or coannihilating with other light sparticles. The first two mechanisms have been tightly constrained by current XENON1T and LHC experiments~\cite{Pozzo:2018anw,Profumo:2017ntc,Abdughani:2017dqs}, while the coannihilation with light sparticles can still be consistent with current data~\cite{Abdughani:2019wss}.

The SUSY contributions to the muon $g-2$ mainly come from the neutralino-smuon and chargino-sneutrino loops. The expressions of one-loop corrections to $a_\mu$ are given by~\cite{Martin:2001st}
\begin{eqnarray}
\delta a_\mu^{\tilde{\chi}^0} & = & \frac{m_\mu}{16\pi^2}
\sum_{i,m}\left\{ -\frac{m_\mu}{ 12 m^2_{\tilde\mu_m}}
(|n^L_{im}|^2+ |n^R_{im}|^2)F^N_1(x_{im})
+\frac{m_{\tilde{\chi}^0_i}}{3 m^2_{\tilde \mu_m}}
{\rm Re}[n^L_{im}n^R_{im}] F^N_2(x_{im})\right\},\phantom{xxxx}\\
\delta a_{\mu}^{\tilde{\chi}^\pm} & = & \frac{m_\mu}{16\pi^2}\sum_k
\left\{ \frac{m_\mu}{ 12 m^2_{\tilde\nu_\mu}}
(|c^L_k|^2+ |c^R_k|^2)F^C_1(x_k)
+\frac{2m_{\tilde{\chi}^\pm_k}}{3m^2_{\tilde\nu_\mu}}
{\rm Re}[ c^L_kc^R_k] F^C_2(x_k)\right\},\phantom{xxxx}
\label{eq:g-2}
\end{eqnarray}
where
\begin{eqnarray}
&& n^R_{im}  =   \sqrt{2} g_1 N_{i1} X_{m2} + y_\mu N_{i3} X_{m1} ,\\
&& n^L_{im}  =   {1\over \sqrt{2}} \left (g_2 N_{i2} + g_1 N_{i1}
               \right ) X_{m1}^* - y_\mu N_{i3} X^*_{m2} ,\\
&& c^R_k  =  y_\mu U_{k2} ,~~~~ c^L_k  =  -g_2V_{k1} ,
\end{eqnarray}
with $i$, $m$ and $k$ being the indices respectively for the neutralinos, smuons and charginos mass eigenstates. $y_\mu = g_2 m_\mu/\sqrt{2} m_W \cos\beta$ being the muon Yukawa coupling. The loop functions $F^N_{1,2}$ and $F^C_{1,2}$, depending on the variables $x_{im}=m^2_{\chi^0_i}/m^2_{\tilde\mu_m}$, $x_k=m^2_{\chi^\pm_k}/m^2_{\tilde\nu_\mu}$, are normalized so that $F^N_{1,2}(1) = F^C_{1,2}(1) = 1$, which can be found in~\cite{Martin:2001st}.
The unitary matrix $X$ that diagonalizes the smuon mass matrix $M^2_{\tilde\mu}$ is given by,
\begin{eqnarray}
X M^2_{\tilde\mu}\, X^\dagger =
{\rm diag}\, (m^2_{\tilde\mu_1}, m^2_{\tilde\mu_2}),
\end{eqnarray}
where
\begin{eqnarray}
M^2_{\tilde\mu}=\left( \begin{array}{cc}
m^2_L +(s_W^2 -\frac{1}{2})m_Z^2\cos 2\beta & m_\mu(A^*_{\mu}-\mu\tan\beta) \\
m_\mu (A_{\mu}-\mu^*\tan\beta) & m^2_R -s_W^2  m_Z^2\cos 2\beta
\end{array}\right),
\end{eqnarray}
in the $\{ \tilde\mu_L, \tilde\mu_R \}$ basis.
Assuming all sparticles have an universal mass $M_{\rm SUSY}$, the SUSY contributions to muon $g-2$ can be approximated as~\cite{Moroi:1996yh}
\begin{eqnarray}
\delta a_\mu^{\rm SUSY} =\frac{\tan\beta}{192\pi^2}
\frac{m_\mu^2}{M_{\rm SUSY}^2}\, (5g_2^2+g_1^2)
= 14 \tan\beta \left ( \frac{100\>{\rm GeV}}{M_{\rm SUSY}} \right )^2 10^{-10} .
\label{msusy}
\end{eqnarray}
It can be seen that the SUSY contributions can be enhanced by a large $\tan\beta$ and suppressed by SUSY mass scale so that heavy SUSY will decouple from such
a low energy observable. To generate sizable contributions to the muon $g-2$, the involved charginos and neutralinos as well as the sleptons cannot be too heavy. From Eq.~\ref{eq:g-2}, we can find that the contribution of chargino-sneutrino loop usually dominates over that of neutralino-slepton loop. But it should be mentioned that a sizable contribution to $g-2$ anomaly can also be from the bino-smuon loop because of the large smuon left-right mixing induced by large $\mu$~\cite{g2-Binosmuon}. 

Two-loop corrections to the muon $g-2$ from fermion/sfermion loops in the MSSM are calculated in~\cite{Fargnoli:2013zia,Fargnoli:2013zda}. These corrections are also significant and even logarithmically enhanced for heavy sfermions. For different masses of sparticles running in the loops, a few percent correction for squark masses in the few TeV region can be obtained. Such a non-decoupling behavior is because that the gaugino and higgs couplings can differ from the corresponding gauge and Yukawa couplings when heavy sfermions are integrated out.

\section{Parameter scan for muon $g-2$ and dark matter}
\label{section 3}
In conjuncture with the requirements of the dark matter relic density and LHC data, we perform our study in the EWSUSY framework, where only electroweakinos and sleptons are light and colored sparticles are heavy. Such a framework allows us to remain agnostic of the detailed UV-physics, yet still capture the features of models for muon $g-2$ and dark matter in the MSSM. We will focus on two promising scenarios: one has bino, winos and sleptons (BWL), and the other has bino, higgsinos and sleptons (BHL). This will narrow down the parameter space of the MSSM and provide a guidance of hunting for electroweakinos and sleptons at the LHC and future colliders. The relevant parameters are scanned in the following ranges:
\begin{eqnarray}
{\rm BWL:} \quad 0{\rm ~TeV} \leq M_1 ,M_2 \leq 3{\rm ~TeV}, ~~3{\rm ~TeV} \leq \mu \leq 5{\rm ~TeV}\\
{\rm BHL:} \quad 0{\rm ~TeV} \leq M_1 ,\mu \leq 3{\rm ~TeV}, ~~3{\rm ~TeV} \leq M_2 \leq 5{\rm ~TeV}
\end{eqnarray}
Other SUSY parameters in both scenarios are taken as
\begin{eqnarray}
 100 {\rm ~GeV} \leq M_{L_{1,2}}=M_{E_{1,2}} \leq 3 {\rm ~TeV}
 \quad 1 \leq \tan\beta \leq 50 \nonumber\\
 3{\rm ~TeV} \leq M_{\widetilde t_R} \leq 5{\rm ~TeV} \quad
-5 {\rm ~TeV} \leq A_t=A_b=A_{\tau} \leq 5 {\rm ~TeV} \nonumber\\
M_{L_3}=M_{E_3}=M_3=5{\rm ~TeV} \quad A_u = A_d=A_e=0
\end{eqnarray}

In our scan we consider the following experimental constraints:
\begin{itemize}
	\item[(1)] We use \textsf{SUSY-HIT}~\cite{susyhit} to calculate the mass spectrum and branching ratios of the particles. We require the Higgs boson $h$ to be SM-like and in the range of $122 < m_H < 128$ {\rm ~GeV}.
	\item[(2)] We impose the constraint of meta-stability of the vacuum state by
                   demanding $|A_t|\lesssim2.67\sqrt{M_{\widetilde{Q}_{3L}}^2+M_{\widetilde{t}_R}^2+M_A^2\cos^2\!\beta}$~\cite{vacuum}.
    \item[(3)] The sleptons must be above 100 GeV, as required by the LEP2 constraints.
	\item[(4)] We calculate the dark matter relic density by \textsf{MicrOMEGAs 4.3.2}~\cite{micromegas} and require its value within $2\sigma$ range of the Planck observed value, $\Omega_{\rm DM}h^2=0.1186\pm0.002$~\cite{planck}.
	\item[(5)] We require the SUSY contribution to explain the current value of muon $g-2$ data $\delta a_{\mu}^{\rm exp-SM}=(2.68\pm0.63\pm0.43)\times10^{-9}$~\cite{gm2-value} within the $2\sigma$ range.                 	
\end{itemize}

\begin{figure}[th]
	\includegraphics[width=16cm,height=8cm]{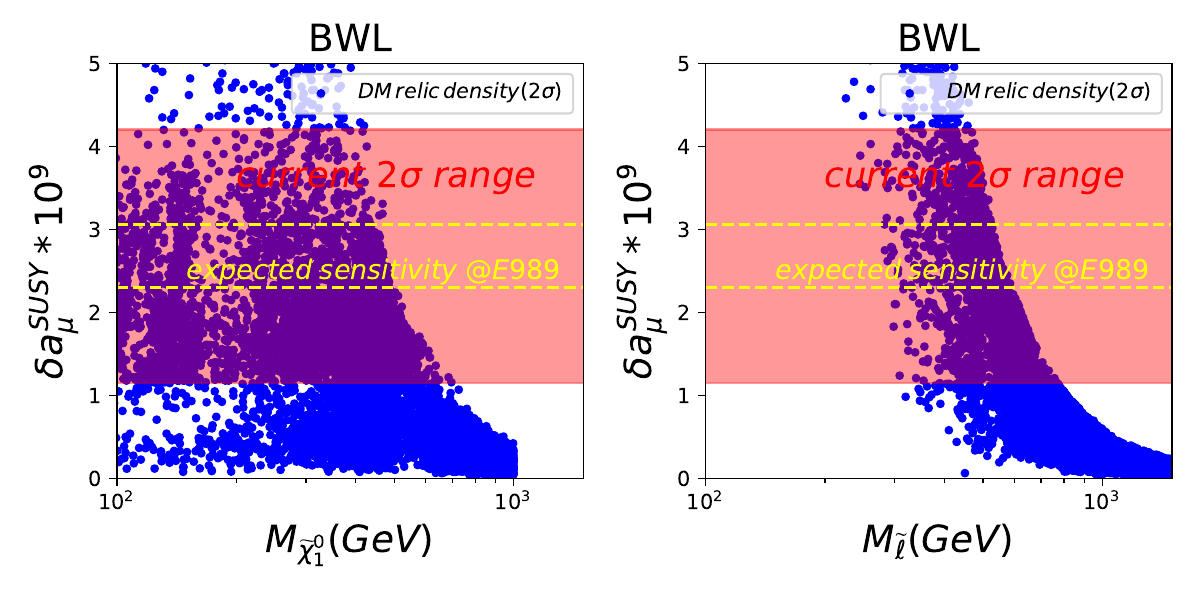}
	\includegraphics[width=16.5cm,height=8cm]{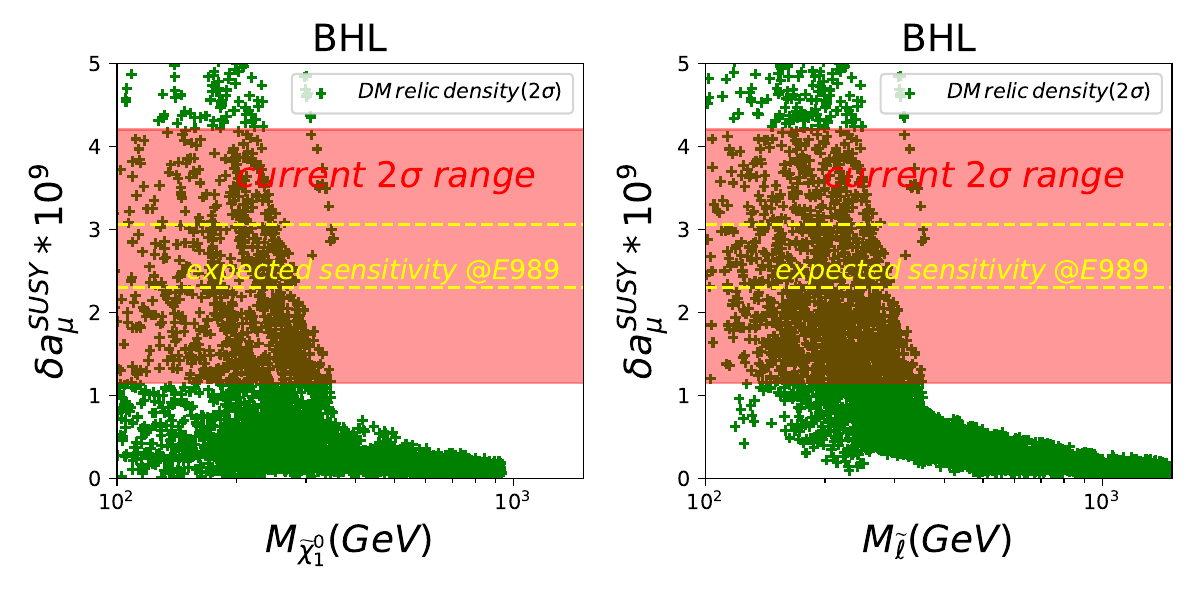}
    \vspace*{-1.0cm}
	\caption{Scatter plots of the two types of samples survived the constraints (1)--(4) on the plane of $\delta a^{\rm SUSY}_\mu$ versus $m_{\tilde{\chi}^0_1}$ and $m_{\tilde{\ell}}$, showing the contributions to the muon $g-2$. The shaded areas are the current $2\sigma$ ranges, while the regions between the dotted lines are the projected $2\sigma$ sensitivity of the experiment at Fermilab (E989), where the expected central value is assumed same as the current experimental value.}
	\label{gm2cut-fig}
\end{figure}
In Fig.~\ref{gm2cut-fig}, we present the contributions of sparticles to the muon $g-2$ for samples survived the constraints (1)--(4) on the plane of $\delta a^{\rm SUSY}_\mu$ versus $m_{\tilde{\chi}^0_1}$ and $m_{\tilde{\ell}}$. The red shaded areas are the $2\sigma$ ranges of explaining current muon $g-2$ anomaly. We find that the dark matter abundance in BWL scenario is achieved mainly through the co-annihilation of the bino-like $\tilde\chi^0_1$ and wino-like $\tilde\chi^0_2$ and $\tilde\chi^\pm_1$. The slepton coannihilation also contributes to the relic density in the relatively heavy mass range. While in the BHL scenario, the correct dark matter abundance is obtained through the co-annihilation of the LSP with sleptons.

In order to interpret the muon $g-2$ deviation, we can see that the masses of $\tilde\chi^0_1$ and $\tilde{\ell}$ have to be lighter than about 700 GeV and 800 GeV in BWL scenario, respectively. But in the BHL scenario, the masses of $\tilde\chi_1^0$ and $\tilde{\ell}$ have to be less than about 350 GeV. On the other hand, we find that the smuon with a mass less than about 200 GeV have been excluded in BWL scenario because of the over-enhancement of $g-2$. On the other hand, a lighter smuon can exist in BHL scenario. This is because that the right-handed smuon in BHL scenario will lead to a negative contribution to $g-2$ so that a lighter smuon is needed to compensate for such a suppression.

By assuming the expected central value same as the current result of $g-2$, we also show the projected $2\sigma$ sensitivity of the E989 experiment at Fermilab that are the regions between the dotted lines. It will further constrain the viable mass ranges of sparticles. To be specific, $\tilde\chi^0_1$ and $\tilde{\ell}$ have to be lighter than about 500 GeV and 600 GeV in BWL scenario, respectively, while in the BHL scenario, the masses of $\tilde\chi_1^0$ and $\tilde{\ell}$ have to be less than about 200 GeV. If these turn out to be true, several popular high scale SUSY models, such as the CMSSM, mSUGRA, GMSB and AMSB, have to be extended because their sfermion spectrum that needs to explain the 125 GeV Higgs mass is too heavy to accommodate muon $g-2$.

\begin{figure}[th]
	\includegraphics[width=16cm,height=10cm]{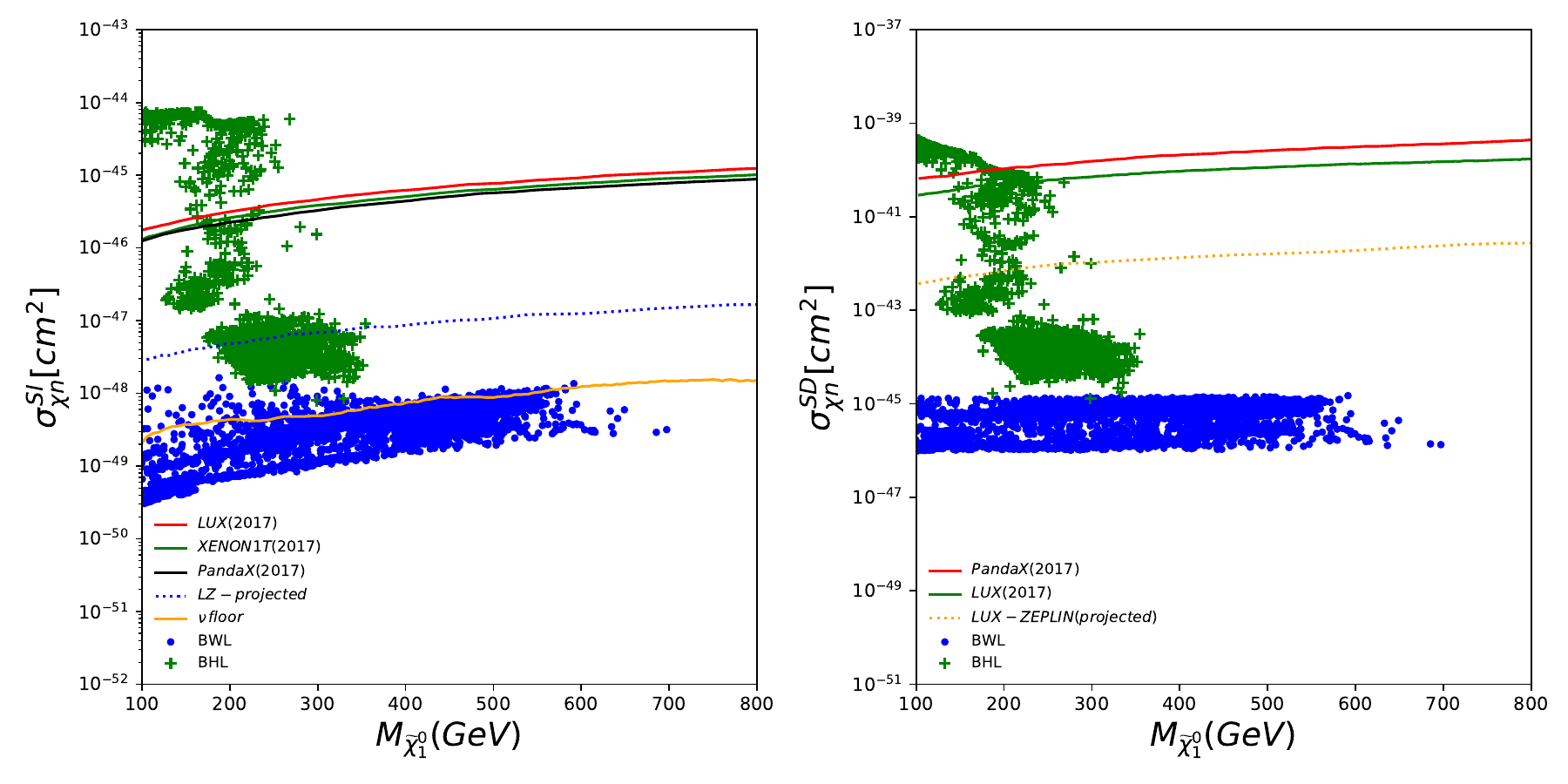}
    \vspace*{-0.5cm}
	\caption{Scatter plots of the samples survived the constraints (1)--(5), showing the spin-independent and spin-dependent neutralino LSP-nucleon scattering cross sections. The samples in BWL and BHL scenarios are denoted by blue dots and green plus respectively. The observed $90\%$ CL upper limits from LUX2017 \cite{LUX2017}, XENON1T-2017 \cite{XENON1T-2017} and PandaX-2017 (Run9+Run10) \cite{PandaX-2017} and the future sensitivities from  LZ-projected \cite{LZ-projected} are shown.}
	\label{SISD-fig}
\end{figure}
In Fig.~\ref{SISD-fig}, we plot the spin-independent and spin-dependent LSP-nucleon scattering cross sections of the samples survived the constraints (1)--(5). Since in BWL scenario the higgsinos are rather heavy and the LSP $\tilde{\chi}^0_1$ is extremely bino-like, it scatters with nucleon very weakly and thus the SI and SD LSP-nucleon scattering cross sections are very small, which can be much below the LZ-projected sensitivities. Those samples may be probed at colliders~\cite{Duan:2018rls,Han:2014xoa}. On the other hand, the LSP $\tilde{\chi}^0_1$ in BHW scenario can have certain higgsino component so that it can scatter with the nucleons sizably and are tightly constrained by current direct detection limits.

\section{observabilities at LHC upgrades and Higgs factory}\label{section 4}

\begin{figure}[ht]
	\includegraphics[width=16cm,height=8cm]{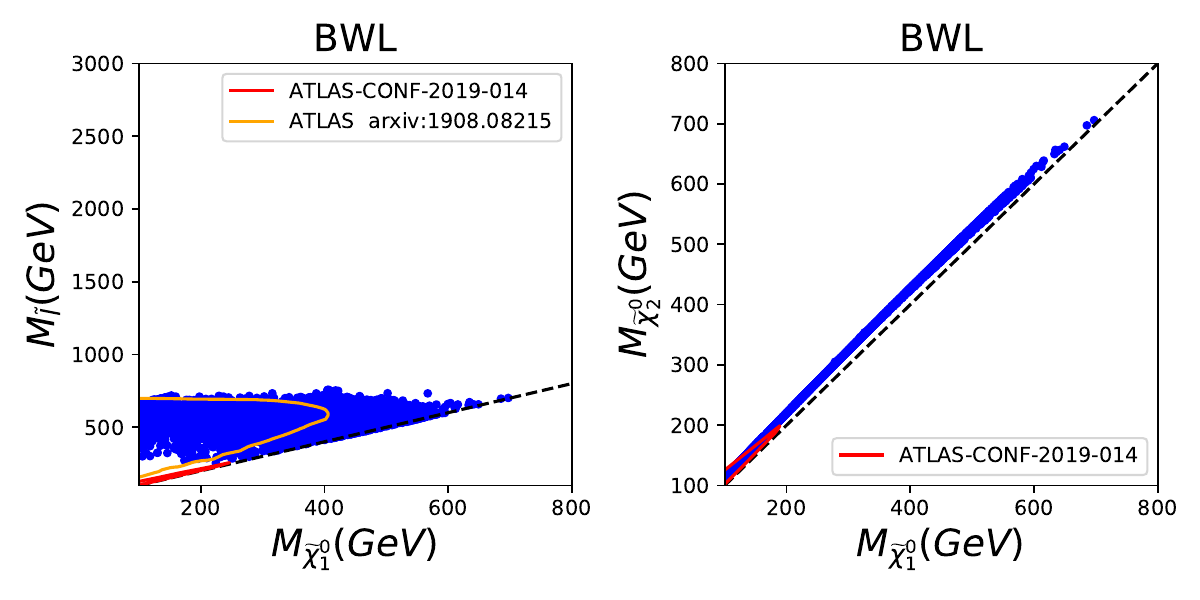}
	\includegraphics[width=16cm,height=8cm]{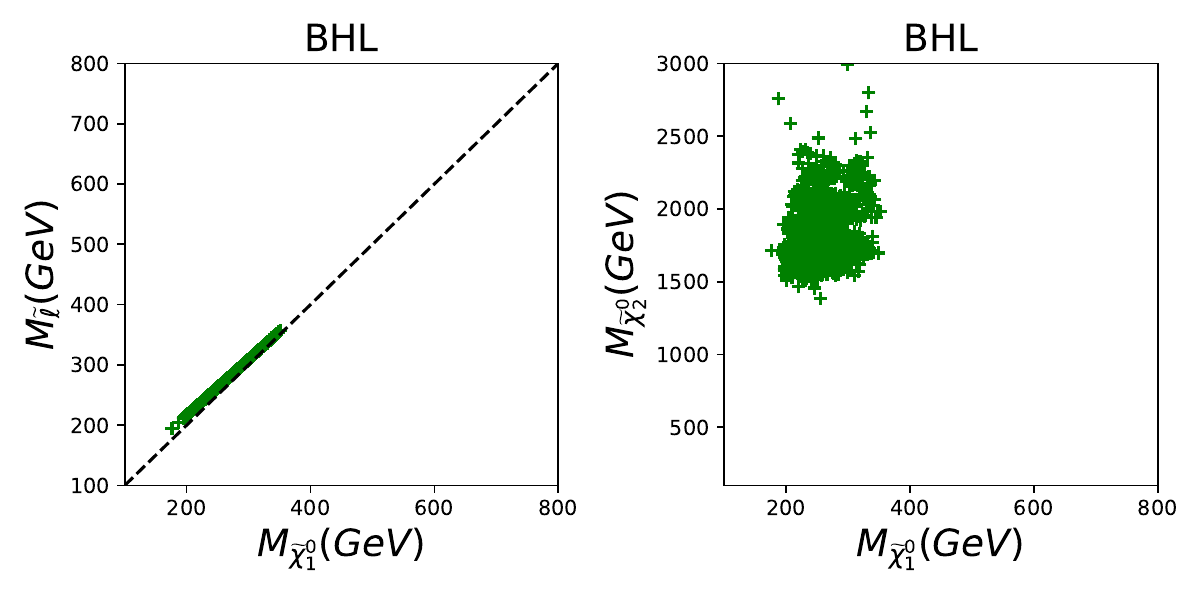}
	\caption{Scatter plots of the samples survived the constraints (1)--(5) and the current direct detection limits,  displayed on the plane of $M_{\tilde{\chi}^0_1}$ versus $M_{\tilde{\chi}^0_2}$ and $M_{\tilde{\ell}}$. The upper and lower panels correspond to the BWL and BHL scenarios, respectively. For the BWL case the regions excluded by ATLAS \cite{atlas-exclude} are shown.}
	\label{M1-fig}
\end{figure}
In Fig.~\ref{M1-fig}, we display the samples survived the constraints (1)--(5) and the dark matter direct detection. The BWL scenario is shown in the upper panel of Fig.~\ref{M1-fig}, where $\tilde\chi^0_1$ is bino-like and $\tilde\chi^0_2$ is wino-like. For most samples the mass difference between the wino-like $\tilde\chi^0_2$ and the bino-like  $\tilde\chi^0_1$ is rather small, while the smuon mass can be quite near or significantly heavier than the mass of $\tilde{\chi}^0_1$.  The BHL scenario is shown in the lower panel of Fig.~\ref{M1-fig}, which has a light spectrum of bino and sleptons but with heavy higgsinos. In this scenario $\tilde\chi^0_1$ is also rather bino-like, albeit with small higgsino admixture, while the $\tilde\chi^0_2$ is higgsino-like. The mass difference between the LSP and sleptons is quite small as well.

We also present the latest exclusion limits from the null results of searching for slepton pair and wino pair at 13 TeV LHC with the luminosity of 139 fb$^{-1}$. For the BWL scenario, a large portion of samples with sizable mass splitting of slepton and LSP have been excluded, which implies a compressed spectrum of bino, wino and sleptons. 
In the meanwhile, the light wino-like $\tilde\chi^0_2$ in the coannihilation region are not allowed either. On the other hand, there is no constraint on the BHL scenario at the LHC because the samples have either heavy higgsino-like $\tilde\chi^0_2$ or compressed slepton and LSP with masses being larger than about 200 GeV.

\begin{figure}[ht!]
	\centering
	\includegraphics[width=16cm,height=6cm]{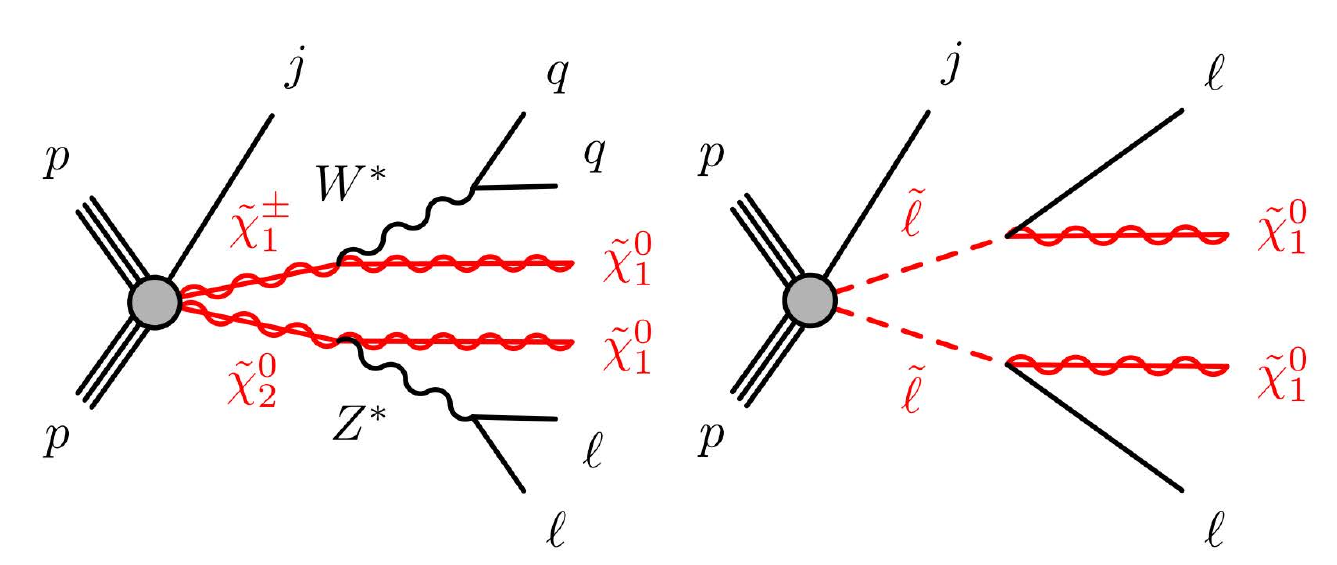}
	\caption{The schematic diagrams of the wino pair production process $pp \to j\tilde{\chi}_2^0\tilde{\chi}^{\pm}_1$ in the BWL scenario (left panel)
   and the slepton pair production process $pp \to j\tilde{\ell} \tilde{\ell}^*$ ($\tilde{\ell}=\tilde{e}_1,\tilde{\mu}_1$) in the BHL scenario (right panel).} 	
	\label{process-fig}
\end{figure}

Next, we investigate the observability of the BWL and BHL scenarios at 27 TeV HE-LHC with 15 ab$^{-1}$ and Higgs factory. Given the production cross section of the wino pair are larger than that of slepton pair, we perform a detailed Monte Carlo simulation of the process $pp\to j\tilde{\chi}^0_2 (\to Z^*\tilde\chi^0_1 \to \ell^+\ell^- \tilde\chi^0_1) \tilde{\chi}^{\pm}_1 (\to W^* \tilde\chi^0_1\to q\bar{q}+\tilde\chi^0_1) \to j +\ell^+\ell^-+\slashed E_T$ for the compressed bino-wino in BWL scenario. While since the sleptons are much lighter than the higgsinos in the BHL, we will analyze the process $pp\to j\tilde{\ell} (\to \ell^- \tilde\chi^0_1) \tilde{\ell}^*(\to \ell^+ \tilde\chi^0_1) \to j+\ell^+\ell^- +\slashed E_T$ for the compressed bino-slepton in the BHL scenario. The schematic diagram of those two process are shown in Fig.~\ref{process-fig}. So in both signal processes, there are a pair of soft opposite-sign same-flavor leptons plus jets plus large missing transverse energy. We will utilize these features to enhance the sensitivity of our signals. The main SM backgrounds come from the Drell-Yan processes, dibosons and the leptonic $t\bar{t}$ events. We generate parton-level events by using \textsf{MadGraph5\_aMC@NLO}~\cite{Madgraph} and then the events are passed to \textsf{Pythia}~\cite{pythia} for showering and hadronization. The detector effects are simulated by \textsf{Delphes}~\cite{delphes}. We perform the analysis of events in the framework of \textsf{CheckMATE2}~\cite{checkmate2}, and evaluate the significance by
\begin{equation}
Z=\frac{S}{\sqrt{S+B+{(\beta B)}^2}},
\end{equation}
where $\beta$ stands for the expected systematic uncertainty. It has to be revisited with the real performance of the upgraded LHC detectors. As a theoretical estimation, we take $\beta=10\%$ in our calculations.

\begin{figure}[ht]
	\centering
	\includegraphics[width=8cm,height=9cm]{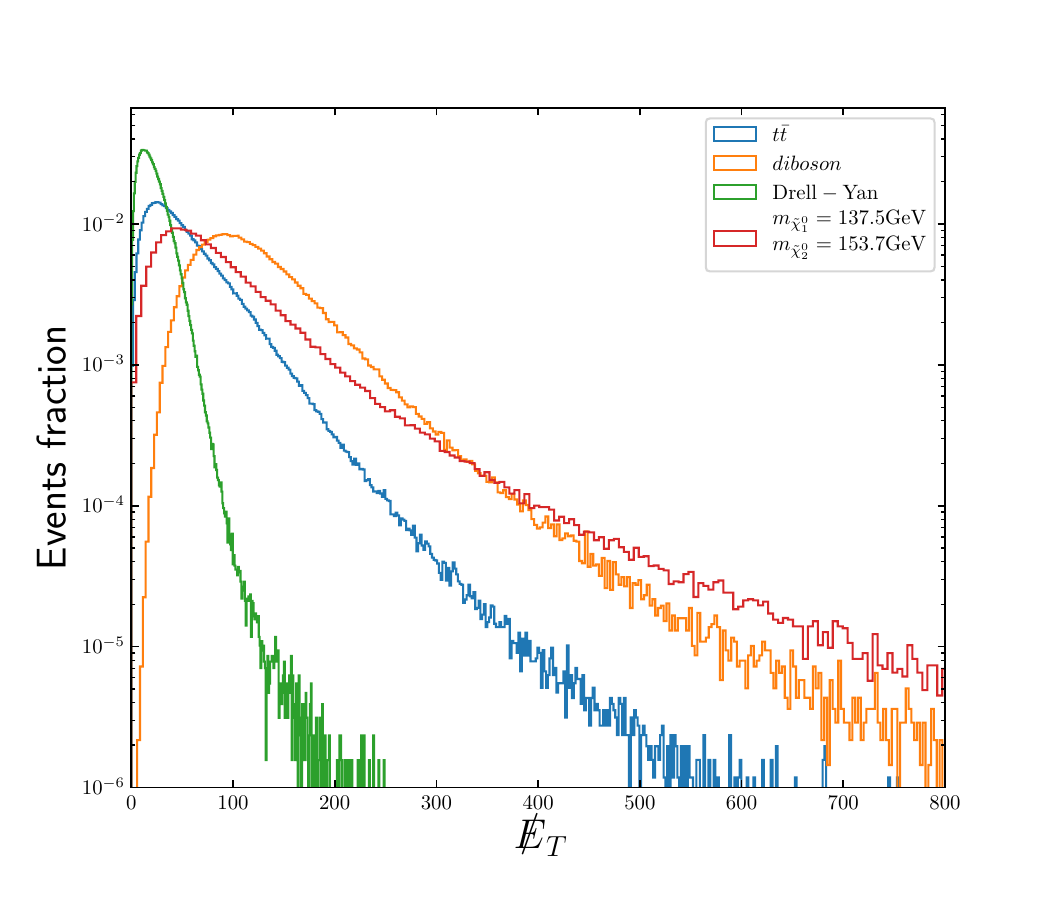}
	\includegraphics[width=8cm,height=9cm]{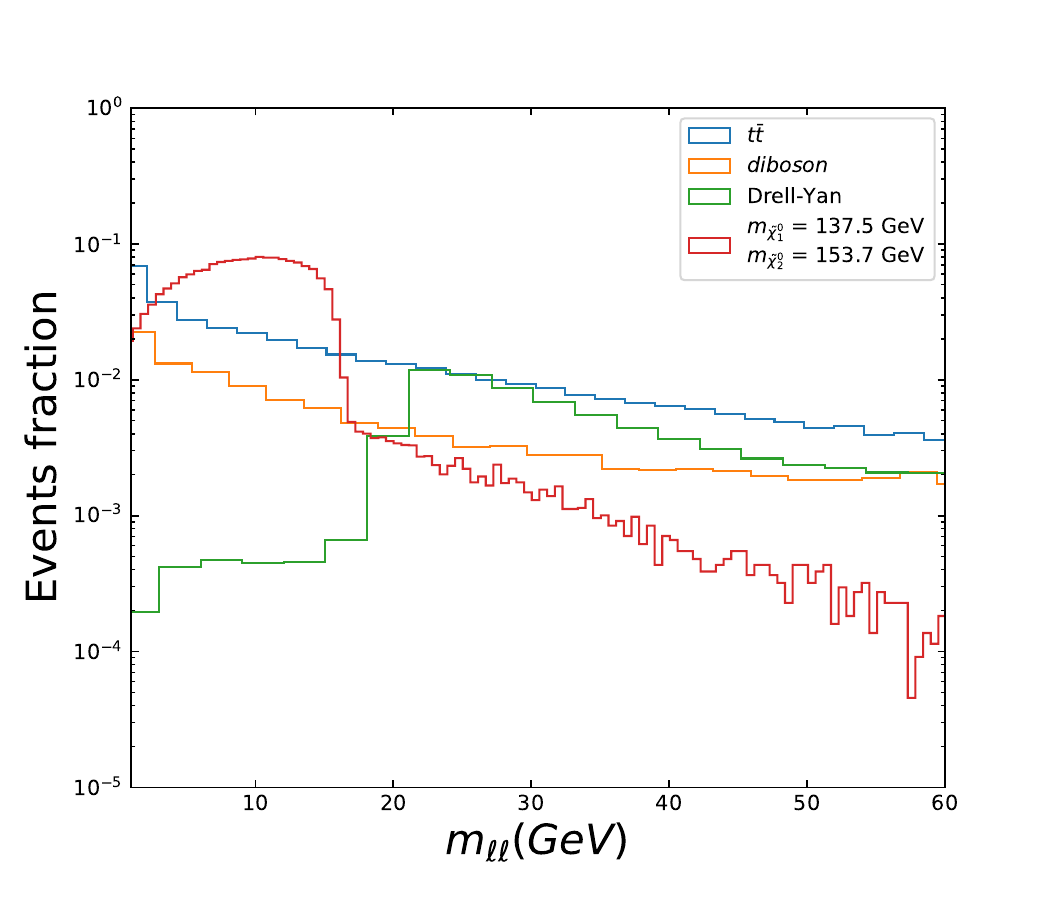}
	\includegraphics[width=8cm,height=9cm]{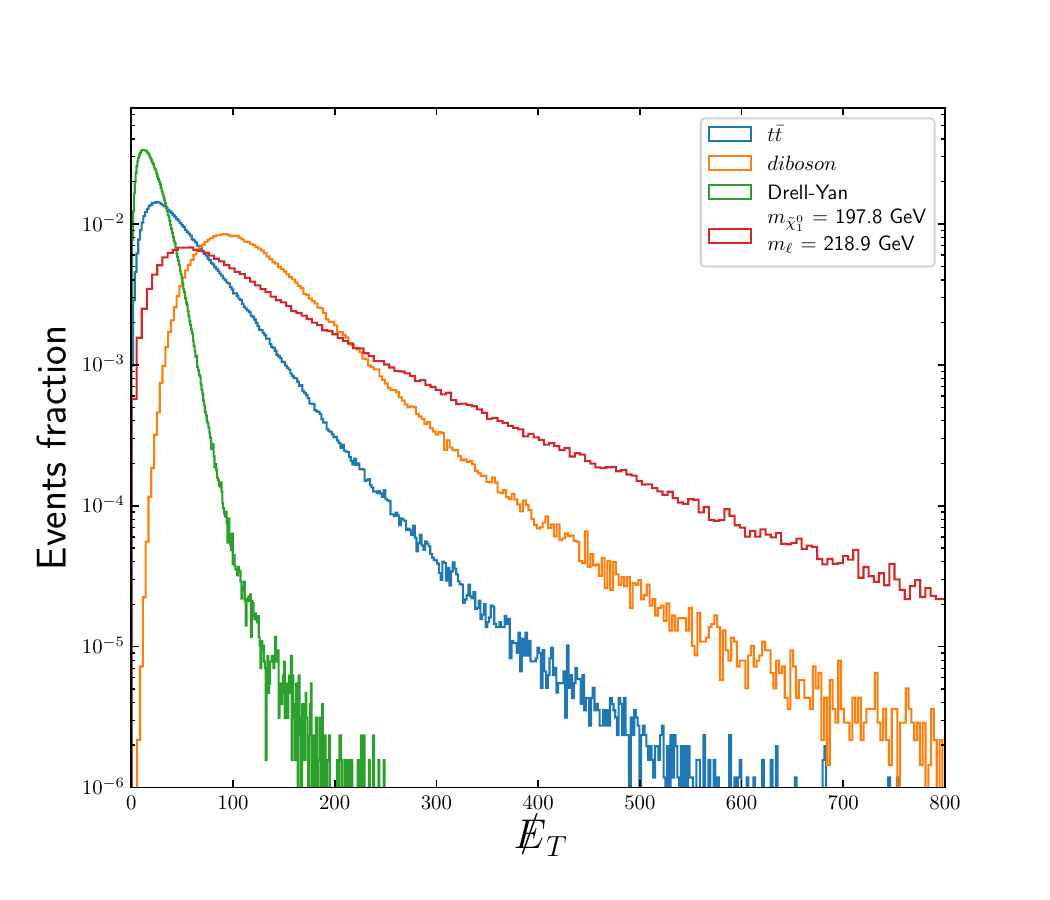}
	\includegraphics[width=8cm,height=9cm]{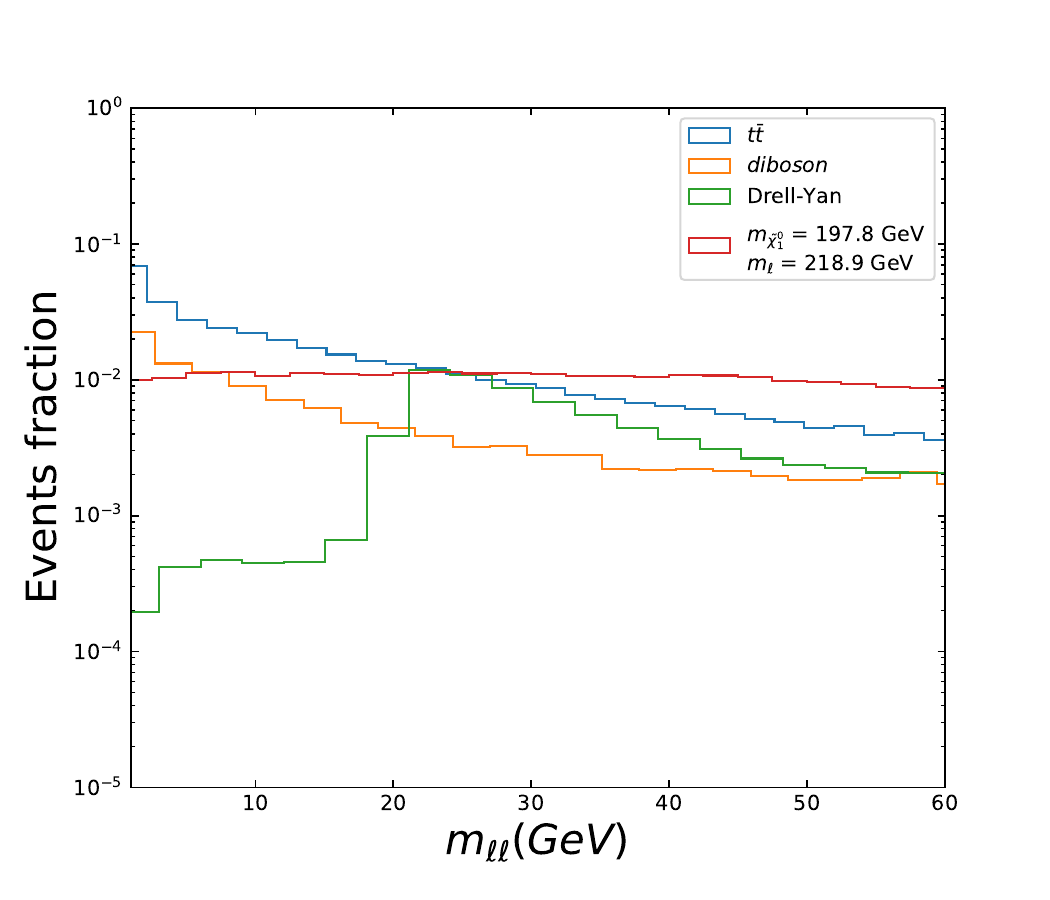}
	\caption{The normalized distributions of $\slashed E_T$ and $m_{\ell \ell}$ of the signal and background events at the 27 TeV HE-LHC. The upper and lower panels are for the BWL and BHL scenarios, respectively.}
	\label{fig:distribution}
\end{figure}

In Fig.~\ref{fig:distribution}, we show the normalized distributions of the missing transverse energy $\slashed E_T$ and the dilepton invariant mass $m_{\ell \ell}$ of the signal and background events. We find that both signals have more events in the range of the large $\slashed E_T$, which can highly suppress the Drell-Yan and $t\bar{t}$ backgrounds. In additional, due to two soft leptons decaying from the sleptons, both signals predict a small value of $m_{\ell \ell}$. According to the kinematical features, we impose the following event selection criteria:
\begin{itemize}
	\item We require the missing transverse energy $\slashed E_T > 200$ GeV.
	\item Two opposite-sign same-flavor (OSSF) leptons are required. The leading and subleading leptons should have the transverse momentum $p_T(\ell_1)>$ 5 GeV and $p_T(\ell_2)>$ 4 GeV. The angular distance $0.05<\Delta R (\ell_1, \ell_2)<2$ in BWL scenario and $0.05<\Delta R (\ell_1, \ell_2)$ in BHL scenario are required.
	\item We require at least one jet and the leading jet $p_T(j_1) > 100$ GeV. The angular separations have to be $\Delta\phi(j_1,P_T^{miss})>2$ and $\Delta\phi(j,P_T^{miss})>0.4$. Also we veto $b$-jets to reduce $t\bar{t}$ background.
	\item We require the dilepton invariant mass 1 GeV $<m_{\ell\ell}<$ 60 GeV and $m_{\ell\ell}\notin(3,3.2)$ GeV to suppress contributions from $J/\psi$ decays and on-shell $Z$ boson decays.
    \item The scalar sum of the lepton transverse momenta $H_{T}^{\rm lep} = p_{T}^{\ell_1} + p_{T}^{\ell_2}$ is small in the compressed region. The ratio $E_{T}^{\rm miss} / H_{T}^{\rm lep}$ can improve the sensitivity for smaller mass splitting. We require $\slashed E_T/H_T^{\rm lep} > \max[5, 15-2m_{\ell\ell}/(1 \rm GeV)]$ for wino pair and $\slashed E_T/H_T^{\rm lep} > \max[3, 15-2[m^{100}_{T_2}/(1 \rm GeV)-100]]$ for slepton pair, where the stransverse mass is defined in \cite{smt}.
    \item The invariant mass $m_{\tau\tau}\notin [0,160)$ GeV can suppress the Drell-Yan background.
\end{itemize}

\begin{table}[h]
	\caption{The cut flow for the cross sections of the signal and backgrounds at the 27 TeV HE-LHC for the BWL benchmark point $m_{\tilde\chi^0_1}=137.4$ GeV, $m_{\tilde\chi^0_2}=m_{\tilde\chi^\pm_1}=153.7$ GeV, $\tan\beta=50$. The cross sections are in units of fb.}
	\label{cutflow-bwl}
	\begin{tabular}{lp{.3cm}rp{.2cm}rp{.2cm}rp{.2cm}r}
		\hline\hline
		Cuts && $t\bar{t}$ && diboson && Drell-Yan&&BWL\\
		\hline\hline
		$\slashed E_T > 200$ GeV && 37512.99&&1721.53 && 246.54 &&618.83  \\
		\hline
		$N(\ell) =2$, OSSF, $p_T(\ell_1) >5$ GeV, $p_T(\ell_2) >4$ GeV && 956.16 && 38.536 && 15.60 &&51.03 \\
		\hline
		$N(j)\geq 1$, $N(b)=0$, $p_T(j_1) > 100$ GeV,\\ $\Delta\phi(j_1,P_T^{\rm miss})>2$, $\Delta\phi(j,P_T^{\rm miss})>0.4$
		&& 74.16 && 16.43  &&9.36  && 34.53 \\
		\hline
		$m_{\tau\tau}\notin [0,160)$ GeV, 1 GeV $<m_{\ell\ell}<60$ GeV,\\$m_{\ell\ell}\notin(3,3.2)$ GeV,  $\Delta R_{\ell\ell} >0.05$
		&& 23.84 && 3.64  && 3.12 && 27.68  \\
		\hline
		$\slashed E_T/H_T^{\rm lep} > \max(5, 15-2m_{\ell\ell})$ &&7.94  && 2.26 && 3.12  &&24.22  \\
		\hline
		$\Delta R_{\ell\ell} <2 $ && 2.65 &&1.38 &&3.12 && 20.15 \\
		\hline
		\hline\hline
	\end{tabular}
	\normalsize
\end{table}

\begin{table}[h]
	\caption{The cut flow for the cross sections of the signal and backgrounds at the 27 TeV HE-LHC for the BHL benchmark point $m_{\tilde\chi^0_1}=197.8$ GeV, $m_{\tilde\ell}=218.9$ GeV, $\tan\beta=17.5$. The cross sections are in units of fb.}
	\label{cutflow-bhl}
	\begin{tabular}{lp{.3cm}rp{.2cm}rp{.2cm}rp{.2cm}r}
		\hline\hline
		Cuts && $t\bar{t}$ && diboson && Drell-Yan&&BHL\\
		\hline\hline
		$\slashed E_T > 200$ GeV && 37512.99 &&1721.53 && 246.54  &&69.97 \\
		\hline
		$N(\ell) =2$, OSSF, $p_T(\ell_1) >5$ GeV, $p_T(\ell_2) >4$ GeV &&956.16  &&38.54  &&15.60  && 7.21\\
		\hline
		$N(j)\geq 1$, $N(b)=0$, $p_T(j_1) > 100$ GeV,\\ $\Delta\phi(j_1,P_T^{\rm miss})>2$, $\Delta\phi(j,P_T^{\rm miss})>0.4$
		&&  74.16 && 16.43 &&9.36  &&4.71 \\
		\hline
		$m_{\tau\tau}\notin [0,160)$ GeV, 1 GeV $<m_{\ell\ell}<60$ GeV,\\$m_{\ell\ell}\notin(3,3.2)$ GeV,  $\Delta R_{\ell\ell} >0.05$
		&& 23.84&&3.64  &&3.12  && 2.18   \\
		\hline
		$\slashed E_T/H_T^{\rm lep} > \max[3, 15-2[m^{100}_{T_2}/(1 \rm GeV)-100]]$ && 13.24 && 3.15 && 3.12  && 1.73  \\
		\hline\hline
	\end{tabular}
	\normalsize
\end{table}

In Tables~\ref{cutflow-bwl} and \ref{cutflow-bhl}, we demonstrate the cut flows for the benchmark points in two scenarios. We can see that the soft OSSF leptons cut will significantly reduce all backgrounds, in particular for $t\bar{t}$ events. The hard $p_T(j_1)>100$ GeV and small dilepton invariant mass 1 GeV $<m_{\ell\ell}<$ 60 GeV can further suppress $t\bar{t}$ and diboson backgrounds by about one order. As pointed in~\cite{atlas-exclude}, the observable $\slashed E_T/H_T^{\rm lep}$ is  useful for reducing the $t\bar{t}$ background.

\begin{figure}[ht]
	\centering
	\includegraphics[width=8cm,height=8cm]{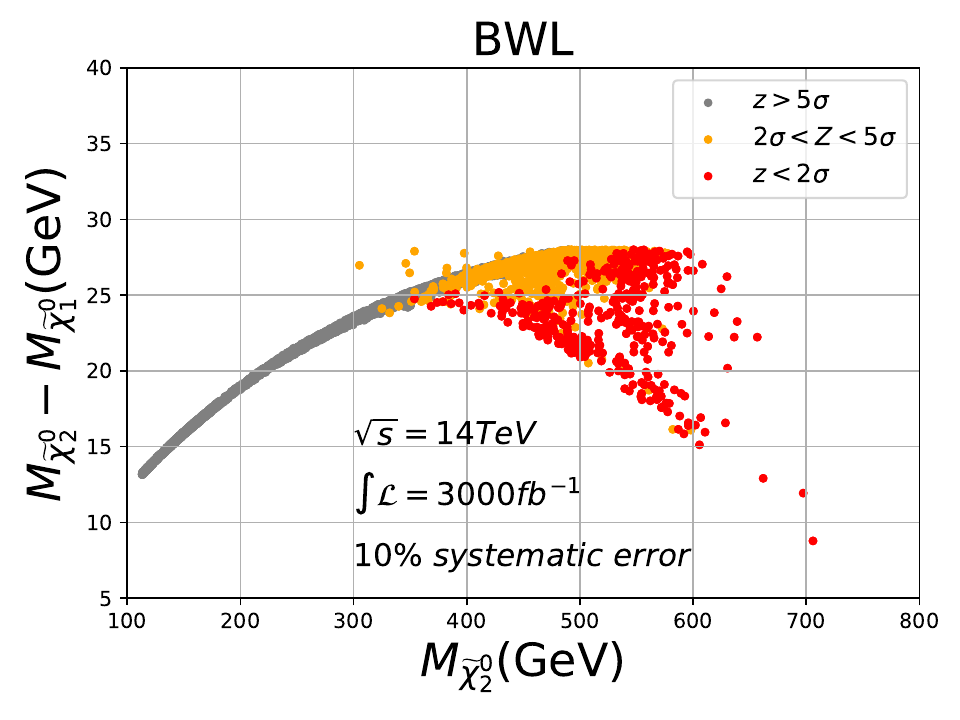}
	\includegraphics[width=8cm,height=8cm]{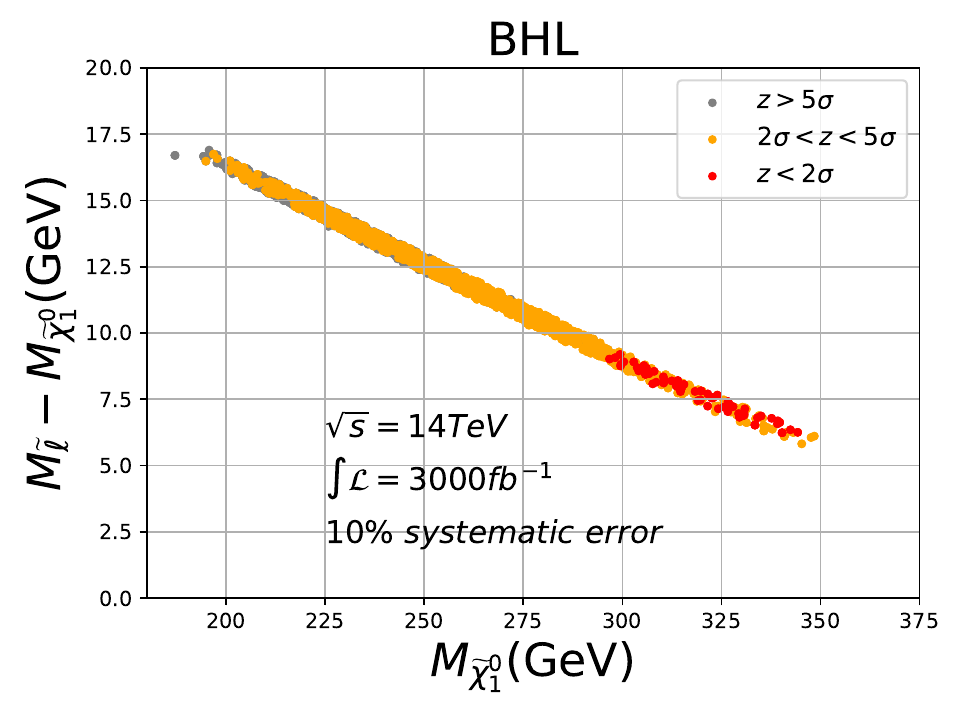}
	\includegraphics[width=8cm,height=8cm]{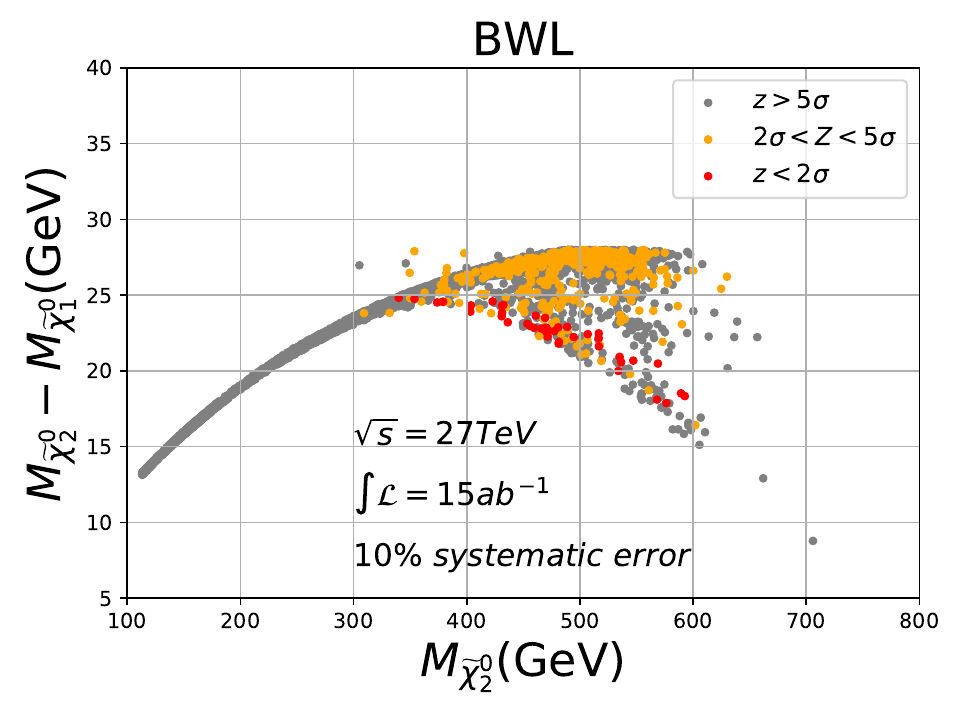}
	\includegraphics[width=8cm,height=8cm]{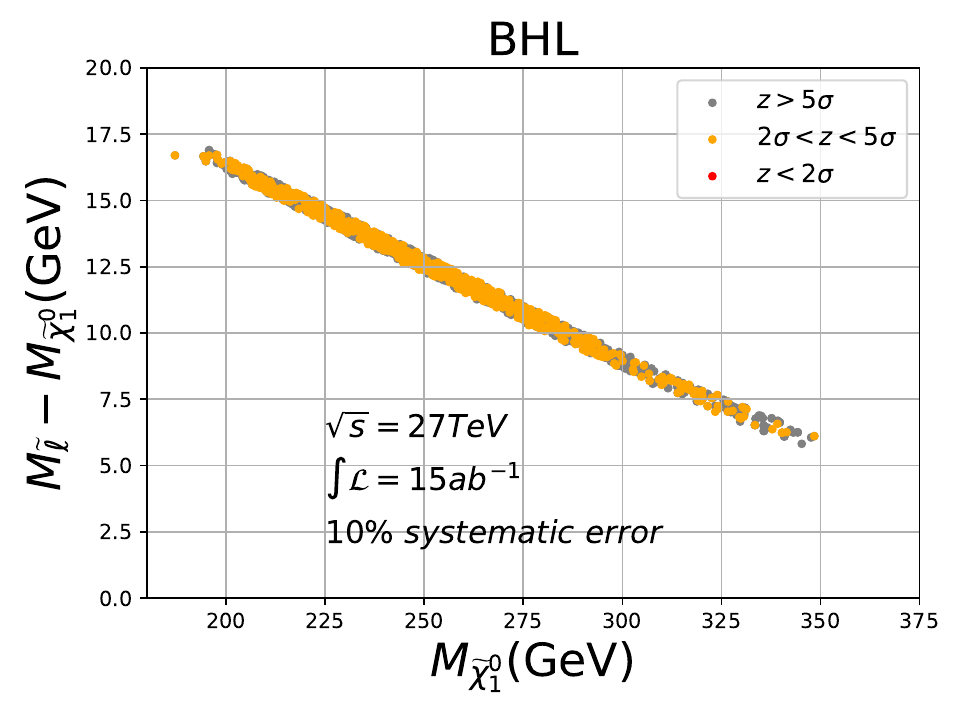}
	\caption{Same as Fig.~\ref{M1-fig}, but showing the significance of the processes  $pp\to \tilde{\chi}_2^0 \tilde{\chi}_1^{\pm} + {\rm jets}$ and $pp\to \tilde{\ell} \tilde{\ell} + {\rm jets}$ at the HL-LHC and HE-LHC. The left and right panels are for the BWL and BHL scenarios, respectively. }
\label{LHC-fig}
\end{figure}
In Fig.~\ref{LHC-fig}, we display the significances of the processes $pp\to j\tilde{\chi}_2^0 \tilde{\chi}_1^{\pm}$ and $pp\to j\tilde{\ell}\tilde{\ell}^*$ at the HL-LHC and HE-LHC. It can be seen that a portion of the samples in both scenarios will be excluded by the search for soft lepton pair plus missing energy events at the HL-LHC. The future HE-LHC is able to further exclude the whole parameter space of BHL and most part of BWL scenarios for satisfying muon $g-2$ and DM experimental results within $2\sigma$ level. We also checked that the 100 TeV proton-proton collider SPPC with the same luminosty cannot do much better than the HE-LHC due to the enhanced backgrounds. On the other hand, it should be mentioned that the heavy higgsinos decaying to light bino in the BHL scenario will provide $3\ell+\slashed E_T$ signature at a 100 TeV hadron collider, which can exclude the higgsino mass up to about 3 TeV at 95\% C.L.. Besides conventional cut-based analysis, the machine learning methods have been recently proposed to enhance the sensitivity in the search of sparticles at the LHC~\cite{Albertsson:2018maf,Abdughani:2019wuv,Abdughani:2018wrw,Ren:2017ymm,Caron:2016hib}. We expect that our result may be improved by using those advanced analysis approaches.

\begin{figure}[ht!]
	\centering
	\includegraphics[width=10cm,height=10cm]{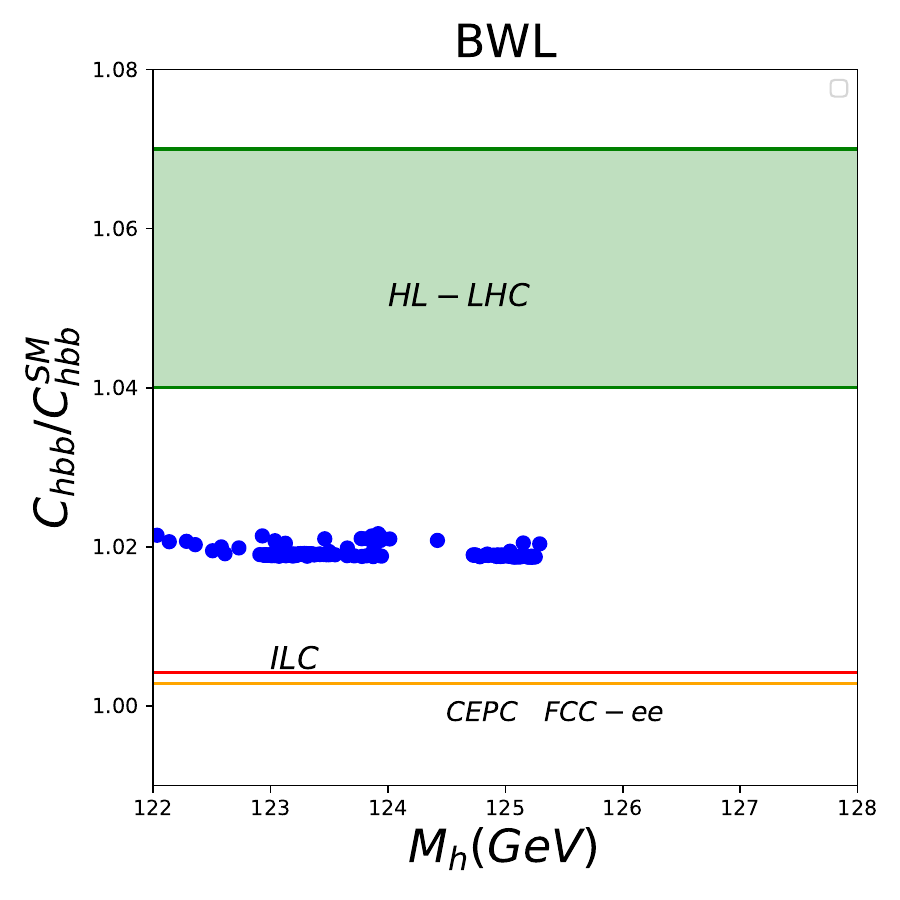}
	\caption{The reduced SM-like Higgs coupling $C_{hb\bar{b}}/C^{SM}_{hb\bar{b}}$ of the samples with the significance $Z<2\sigma$ for the BWL scenario in Fig.~\ref{LHC-fig}. The sensitivities of the HL-LHC (14 TeV, 3 ab$^{-1}$)~\cite{HL-LHC}, ILC (250 GeV, 2 ab$^{-1}$)~\cite{ILC}, FCC-ee (240 GeV, 5 ab$^{-1}$)~\cite{FCC} and CEPC (240 GeV, 5 ab$^{-1}$)~\cite{CEPC} to the Higgs couplings are also shown.}
	\label{hcoupling-fig}
\end{figure}

Since the LHC experiment has been continuously pushing up the new physics scale, the future $e^+e^-$ Higgs factory, either CEPC, FCC-ee or ILC, has
limited energy to directly produce new particles. However, due to its clean environment, such a Higgs factory is a precision test machine and can measure the Higgs couplings at one percent level or better, which may reveal the new physics effects through the Higgs couplings~(see examples, \cite{Cao:2013ur,Hu:2014eia,Cao:2014rma,Kobakhidze:2016mfx,Wu:2019hso,Han:2014qia}).

As shown in the above section, the electroweakinos and sleptons cannot be too heavy in order to explain the muon $g-2$ and provide the correct dark matter abundance. These light uncolored SUSY particles may cause some indirect effects in the Higgs couplings. Among the Higgs couplings, the $hb\bar{b}$ and $h\tau^+\tau^-$ couplings can still deviate from the SM predictions sizably~\cite{Wu:2015nba}. In the following we demonstrate the $hb\bar{b}$ coupling as an illustration.

In Fig.~\ref{hcoupling-fig}, we display the $hb\bar{b}$ coupling for the samples in Fig.~\ref{LHC-fig} that cannot be excluded at the HE-LHC. At tree level the $hb\bar{b}$ coupling is given by $g(m_b/2m_W)(\sin\alpha/\!\cos\beta)$ and the one-loop corrections are presented in \cite{Carena:1995bx}. In our calculations we use the package \textsf{FeynHiggs-2.11.3} \cite{feynhiggs} which includes the one-loop effects and also various two-loop contributions.
The sensitivities of the HL-LHC (14 TeV, 3 ab$^{-1}$), ILC (250 GeV, 2 ab$^{-1}$), FCC-ee (240 GeV, 5 ab$^{-1}$) and CEPC (240 GeV, 5 ab$^{-1}$) to the Higgs couplings are also shown. We can see that in the BWL scenario the $hb\bar{b}$ coupling can still be enhanced by about two percent, which is below the HL-LHC sensitivity but can be readily covered by the Higgs factory ILC, FCC-ee, or CEPC. Therefore, the precision measurement of the Higgs couplings could play a complementary role of probing such a scenario at future high energy lepton collider.

\section{Conclusions}\label{section5}
Since the colored sparticles have been excluded up to TeV scale, searching for the electroweak supersymmetry will be one of the major tasks in future experiments. Besides the LHC, the on-going muon $g-2$ and dark matter experiments provide another good place to hunt for electroweakinos and sleptons in EWSUSY. In this work, we examined the parameter space of EWSUSY under the constraints of the muon $g-2$ anomaly, the DM relic density, the DM direct detections and the LHC data. By analyzing the survived samples, we obtained the following observations:
(1) There are two viable scenarios for explaining the muon $g-2$ anomaly. One has the light compressed bino, winos and sleptons (BWL), and the other has light compressed bino and sleptons but heavy higgsinos (BHL). In the BHL scenario, the masses of $\tilde\chi_1^0$ and $\tilde{\ell}$ have to be smaller than about 350 GeV. In the BWL scenario, the masses of $\tilde\chi^0_1$ and $\tilde{\ell}$ have to be smaller than about 700 GeV and 800 GeV, respectively. If this anomaly persists in the on-going E989 experiment, the allowed parameter space will be further narrowed.
(2) In both scenarios, the dark matter has to be the bino-like neutralino and the dominant annihilation mechanism to achieve the correct dark matter abundance is through the bino-wino or bino-slepton coannihilation. Also, we found that the sneutrino DM or the neutralino DM with sizable higgsino component has been excluded by direct detections, due to the large scattering cross section of dark matter and nucleus.
(3) The BWL scenario has been tightly constrained by the latest LHC Run-2 results of searches for soft $\ell^+\ell^- + \slashed E_T$ events from slepton pair, which implies a compressed spectrum of bino, winos and sleptons. In contrast, the BHL scenario can escape the current LHC limits. We explored the observability of these sparticles in both scenarios at future colliders. We found that the HE-LHC with the luminosity $L=15$ ab$^{-1}$ can exclude the whole BHL scenario and most part of BWL scenarios at $2\sigma$ level. The rest of samples that alter the Higgs coupling by two percent level may be excluded by the precision measurement of the Higgs couplings at a future Higgs factory.

\section*{Acknowledgments}
    We thank Chengcheng Han and Jie Ren for useful discussions. Part of this work was done while M. A. was visiting Nanjing Normal University.
 	This work was supported by the National Natural Science Foundation of China (NNSFC) under
	grant Nos.~11705093, 11675242, 11821505, and 11851303,
	by Peng-Huan-Wu Theoretical Physics Innovation Center (11847612),
	by the CAS Center for Excellence in Particle Physics (CCEPP),
	by the CAS Key Research Program of Frontier Sciences
	and by a Key R\&D Program of Ministry of Science and
	Technology under number 2017YFA0402200-04.

\end{document}